\documentclass[fleqn,usenatbib,useAMS]{mnras}

\usepackage[T1]{fontenc}
\usepackage{ae,aecompl}

\usepackage[dvipdfmx]{graphicx}	
\usepackage{amsmath}	
\usepackage{amssymb}	
\usepackage{mathptmx}
\usepackage{ulem}

\newcommand{\p}{Pop. III }

\newcommand{\MO}{\, M_\odot}
\newcommand{\hMpc}{$ \, h^{-1} \rm Mpc$}
\newcommand{\hpc}{$ \, h^{-1} \rm pc$}

\newcommand{\Dmax}{D_{\rm max}}
\newcommand{\Dmin}{D_{\rm min}}
\newcommand{\dNacc}{\dot{N}_{\rm acc}}
\newcommand{\dNaccz}{\dot{N}_{{\rm acc}, 0}}
\newcommand{\dMacc}{\dot{M}_{\rm acc}}
\newcommand{\dMaccz}{\dot{M}_{{\rm acc}, 0}}
\newcommand{\kms}{{\rm km~s^{-1}}}

\title[Metal pollution of Pop. III survivors via ISOs]{Effect of interstellar objects on metallicity of low-mass first stars formed in a cosmological model}
\author[]{
Takanobu Kirihara,$^{1,2}$\thanks{E-mail: tkirihara@chiba-u.jp}
Ataru Tanikawa$^{3,4}$ and
Tomoaki Ishiyama$^{1}$
\\
$^{1}$Institute of Management and Information Technologies,
 Chiba University, 1-33, Yayoi-cho, Inage-ku, Chiba, 263-8522, Japan\\
$^{2}$Department of Physics, Graduate School of Science,
 Chiba University, 1-33, Yayoi-cho, Inage-ku, Chiba, 263-8522, Japan\\
$^{3}$Department of Earth Science and Astronomy, College of
 Arts and Sciences, The University of Tokyo, 3-8-1 Komaba, Meguro-ku,
 Tokyo 153-8902, Japan\\
$^{4}$RIKEN Advanced Institute for Computational Science,
 7-1-26 Minatojima-minami-machi, Chuo-ku, Kobe, Hyogo 650-0047, Japan
}

\date{Accepted 2019 May 7. Received 2019 April 29; in original form 2019 February 18}

\pubyear{2019}

\begin{document}
\label{firstpage}
\pagerange{\pageref{firstpage}--\pageref{lastpage}}
\maketitle

\begin{abstract}

We investigate metal pollution onto the surface of low-mass population III stars (\p survivors) via interstellar objects floating in the Galactic interstellar medium.
Only recently, Tanikawa et al. analytically estimated how much metal should collide to an orbiting \p survivor encouraged by the recent discovery of 'Oumuamua and suggested that ISOs are the most dominant contributor of metal enrichment of \p survivors. 
When we consider a distribution of interstellar objects in the Galactic disc, \p survivors' orbits are significant properties to estimate the accretion rate of them though Tanikawa et al. assumed one modelled orbit. 
To take more realistic orbits into calculating the accretion rate, we use a high-resolution cosmological $N$-body simulation that resolves dark matter minihaloes. 
\p survivors located at solar neighbourhood have a number of chances of ISO($> 100$~m) collisions, typically $5\times10^6$ times in the last $5$~Gyr, which is one order of magnitude greater than estimated in the previous study. 
When we assume a power-law parameter $\alpha$ of the ISO cumulative number density with size greater than $D$ as $n \propto D^{-\alpha}$, $0.80 \MO$ stars should be typically polluted [Fe/H]$\sim -2$ for the case of $\alpha=2.0$. 
Even in the cases of $0.70$ and $0.75 \MO$ stars, the typical surface metallicity are around [Fe/H]$=-6 \sim -5$. 
From the presence of stars with their [Fe/H], we can constrain on the lower limit of the power $\alpha$, as $\alpha \gtrsim 2.0$, which is consistent with $\alpha$ of km-size asteroids and comets in the solar system. 
Furthermore, we provide six candidates as the ISO-polluted \p stars in the case of $\alpha \sim 2.5$. 
Metal-poor stars so far discovered are possible to be metal-free \p stars on birth. 

\end{abstract}



\section{Introduction}

Formation of the first stars, so-called Population III (Pop. III) stars, is expected to play an essential role in cosmic history. 
\p stars were formed in pristine gas and marked the end of ``Dark Age'' of the universe. 
They produced first heavy elements in the universe, and the metals derive the next generation stars. 
To investigate the nature of the first objects, a number of studies have simulated the formation of primordial clouds considering H$_2$ cooling \citep[e.g.,][]{Tegmark1997, Omukai1998, Omukai2001, Omukai2003, Nakamura2001, Abel2002, Schneider2002, Bromm2002, Yoshida2006, Yoshida2008, Bromm2009, Hosokawa2011, Stacy2012}. 
In the Lambda cold dark matter universe, the formation of \p stars began at $z\sim 20$ in dark matter minihaloes with their mass of $10^5-10^6 \MO$ \citep{Abel2002, Bromm2002, Bromm2009}. 
Cosmological radiation hydrodynamical simulations have suggested that typical \p stars are very massive $10-1000 \MO$ \citep{Susa2014, Hirano2014, Hirano2015}. 
\p star formation in minihaloes lasts in reionisation epoch ($z\sim10$) due to the photodissociation of H$_2$ molecule. 
Since lifetimes of such massive stars are short $\sim 10$~Myr, in situ \p stars should be explored in the high-redshift universe. 
Therefore, direct observation of the signal is still quite difficult \citep{Barkana2018}. 

Alternatively, we have a chance to directly explore \p stars in the Milky Way (MW). 
If \p stars are born as low-mass stars and their mass is $\lesssim 0.8 \MO$, they have much longer lifetimes and could survive in the present-day universe (hereafter, \p survivors). 
As recent cosmological hydrodynamic simulations show, circumstellar discs around first protostars are unstable and gravitational fragmentation is induced \citep[e.g.,][]{Machida2008,Clark2011b, Clark2011, Greif2011, Greif2012, Susa2014}. 
Low-mass \p stars are expected to be born in the discs via such fragmentation \citep[e.g.,][]{Greif2012, Susa2019}. 
However, no metal-free star has been discovered so far although there have been great efforts for finding \p survivors with large surveys of more than $10^5$ field stars in the MW \citep[][and the references therein]{Frebel2015}. 

One possible interpretation of the lack of metal-free stars in the MW is that metals in the interstellar medium might pollute their surface \citep{Yoshii1981, Komiya2009, Komiya2015, Shen2017}. 
It means that \p survivors might not be able to keep metal-free in the MW. 
\citet{Komiya2015} showed the surface of \p stars can be polluted up to [Fe/H]$\sim -5$. 
\citet{Shen2017} investigated such metal pollution using a cosmological simulation and showed the typical metallicity of \p stars with $0.8 \MO$ is [Fe/H]$=-6\sim-5$. 
In the context, metal enrichment occurs in the shallow convection layer; therefore, a relatively small amount of accreted heavy elements affects the observed metal abundance of stars. 
When calculating gas accretion, however, Bondi-Hoyle accretion model was always adopted \citep{Komiya2009, Komiya2015, Shen2017}. 
\citet{Tanaka2017} and \citet{Suzuki2018} evaluated the effects of stellar wind to such accreting gas and concluded that stellar wind prevents accretion of heavy elements. 
Then, the resultant metallicity becomes only up to [Fe/H]$\sim -14$ \citep{Tanaka2017}. 
Interstellar dust should not pollute \p survivors, either. 
It is sublimated to gas by stellar radiation before it reaches to stellar surface, and is blown away by stellar wind. 

Only recently, \citet{Tanikawa2018} first pointed out the importance of the metal pollution of colliding interstellar objects (ISOs) or interstellar asteroids like 'Oumuamua, which is the first ISO observed passing through the Solar System using the Pan-STARRS1 telescope \citep{Meech2017}. 
\citet{Tanikawa2018} analytically estimated how much metal should collide to an orbiting \p survivor. 
In their study, they focused on how ISOs provided their metal varying the size distribution of ISOs and suggested that ISOs are the most dominant contributor of metal enrichment of \p survivors. 

After the discovery of 'Oumuamua, the origin and the number density of similar objects have been actively discussed. 
\citet{Do2018} estimated the number density of ISOs as 0.2~${\rm au}^{-3}$ considering the detection of 'Oumuamua with the survey volume and time of Pan-STARRS. 
They considered the origin as the released object of exoplanetary Oort cloud at the end of a star's main-sequence lifetime and the progenitor might be a white dwarf. 
Similar objects could be produced in tidal disruption events of asteroids or planets by white dwarfs \citep{Rafikov2018}, and in a fragment of a comet-like planetesimal born in a planet-forming disc \citep{Raymond2018}.
The high number density can be explained by the amount of debris ejected during the formation process of planets \citep{Zwart2018}. 
However, the situation that non-gravitational acceleration was observed despite no sign of a cometary tail \citep{Meech2017,Fitzsimmons2018} is still puzzling \citep{Bialy2018}. 

In this paper, we revisit investigating metal pollution onto \p survivors via ISOs floating in the Galactic interstellar medium. 
One key parameter is the accretion rate of ISOs that would be significantly affected by the orbit of \p survivors because how long the survivors spend their lives in ISO-rich region decides it. 
However, the estimation of the accretion rate in \citet{Tanikawa2018} was derived by assuming one modelled survivor's orbit. 
To take more realistic orbits into calculating the accretion rate, we use a high-resolution cosmological simulation and a \p formation model carried out by \citet{Ishiyama2016}. 
The cosmological $N$-body simulation resolves dark matter minihaloes to predict where low mass \p survivors are distributed in the halo of the MW.

This paper is organised as follows. 
In Section~\ref{sec:numerical}, we describe our numerical modelling for the estimation of metal pollution of \p survivors using the cosmological $N$-body simulation. 
The accretion rate of ISOs to \p stars and the resultant surface metallicity of \p stars are described in Section~\ref{sec:ISOacc}. 
In Section~\ref{sec:Discussion}, we state discussions for recent observations of metal-poor stars. 
In Section~\ref{sec:summary}, we present a brief summary. 

\section{Numerical modelling}
\label{sec:numerical}
\subsection{Cosmological $N$-body simulation}

We focus on orbits of surviving low-mass \p stars in the MW via hierarchical structure formation based on the Lambda cold dark matter model. 
For this purpose, we adopt a high-resolution cosmological $N$-body simulation conducted by \citet{Ishiyama2016} using a massively parallel TreePM code GreeM \citep{Ishiyama2009,Ishiyama2012}. 
The simulation adopts the cosmological parameter set of $\Omega_0=0.31$, $\Omega_b=0.048$, $\lambda_0=0.69$, $h=0.68$, $n_s=0.96$ and $\sigma_8=0.83$ \citep{Planck2014}, and is run with $2048^3$ dark matter particles in a comoving box of $8$~\hMpc. 
A particle mass is $5.13\times10^3\, h^{-1}\, M_{\odot}$ and the gravitational softening is $120$~\hpc. 

It is widely believed that \p stars are born in H$_2$ molecule cooling minihaloes without metals. 
\citet{Ishiyama2016} combined the cosmological $N$-body simulation and the \p star formation modelling the \p formation under cosmic UV radiation. 
They imposed the upper and lower criteria of the virial temperature of the minihaloes for the \p formation. 
Dark matter haloes are identified by Friends-of-Friends (FoF) algorithm \citep{Davis1985} adopting a linking parameter of $b=0.2$. 
The details of constructing merger trees are described in \citet{Ishiyama2015}. 
The smallest haloes consist of $32$ particles that resolve dark matter minihaloes where \p stars are expected to be formed. 
In this study, we assume that one low-mass ($0.7-0.8 \MO$) \p survivor exists in each minihalo, and the star is represented by the particle which is gravitationally most bound at the formation redshift of the \p star. 

The cosmological simulation contains nine MW-sized haloes at $z=0$. 
Only to obtain time evolution of MW-sized haloes, we use catalogues generated by ROCKSTAR (Robust Overdensity Calculation using K-Space Topologically Adaptive Refinement) halo finder \citep{Behroozi2013a} and a consistent trees code \citep{Behroozi2013b}. 
In post-process, we track \p survivors' orbits and calculate how long and where they are in the disc. 
We track \p survivors' orbits adopting a linear interpolate between snapshots of the cosmological $N$-body simulation spaced by 0.01 in $\Delta {\rm log}_{10}(1+z)$, and the number of interpolation is set to 100. 
For the purpose, we embed a stable Galactic disc to the central region of the host halo taken from the cosmological $N$-body simulation. 
As a working hypothesis, the angular momentum of the disc is always set to point the same direction as that of the host halo. 
For a safety selection, we select a halo which does not experience a major merger since 5~Gyr ago and whose angular momentum changes little since then. 
\citet{Sales2012} suggested that disc galaxies tend to have roughly aligned angular momentum from inner region to outer region. 
The selected halo is the third most massive halo, which has the virial mass and virial radius of $2.27\times10^{12} \MO$ and 345~kpc, respectively. 
We rotate all the halo particles as the pole of the disc points z-axis in cartesian coordinate.

\subsection{ISO number density and size distribution}
\label{sec:ISONumberDensity}

ISOs are made from heavy elements. 
The spatial distribution of ISOs is still under debate, however, one possibility of their origin is that planet-forming discs around Pop. I stars produce them. 
Therefore, ISOs are thought to be distributed in the Galactic disc, which consists of metal-rich Pop. I stars \citep{Seligman2018}. 
To calculate the ISO accretion rate, we assume that the spatial distribution of ISOs is proportional to the stellar density of the Galactic thin disc. 

We evaluate the ISO accretion rate assuming two disc models (a) exponential disc model and (b) flat disc model.  
The flat disc model is a simple thin disc model mimicking \citet{Tanikawa2018}, which has the thickness and the rotation velocity of 400~pc and 220~$\kms$, respectively. 
We adopt the exponential disc, which can be expressed as
\begin{equation}
\Sigma_{\rm d}(R)=\Sigma_0\;{\rm exp}\left(-\frac{R}{a_{\rm d}}\right),
\end{equation}
where $\Sigma_0$ and $a_{\rm d}$ are the central surface brightness and the scale radius of the disc, respectively.
\cite{Sofue2016} fit the rotation velocity of the MW assuming multiple components such as the central black hole, a bulge, an exponential flat disc and a dark matter halo, and obtained $\Sigma_0=6.0\times 10^{8}$~$M_{\odot}\, {\rm kpc}^{-2}$ and $a_{\rm d}=4.9$~kpc. 
Assuming the thickness of the disc of 400~pc, which is used by \citet{Tanikawa2018}, we convert the surface density profile of the disc to the spatial distribution of ISOs. 
The vertical number density of ISOs is set to be constant at a Galactocentric radius in the disc plane. 
We consider a cumulative number density $n_0$ for ISOs with their radius larger than $D_0$. 
We set $n_0=0.2$~au$^{-3}$ at the solar neighbourhood adopting an estimation from the observation of `Oumuamua and the Pan-STARRS survey volume \citep{Do2018}. 
The radial number density of ISOs in the exponential disc is normalised to $n_0$ at the solar neighbourhood. 
Then we obtained the spatial cumulative number density distribution of the ISOs as 
\begin{equation}
n_0(R)=n_0\, {\rm exp}\left(\frac{8\,{\rm kpc}-R}{a_{\rm d}}\right). \label{eq:galactic_n_ISOs}
\end{equation}
The motion of ISOs is assumed to be synchronised with the rotation of the disc. 
We set the edge of the disc models to the Galactocentric distance of $10$~kpc. 

In this study, we adopt a power-law distribution of the ISO size distribution following \citet{Tanikawa2018}. 
We give the cumulative number density of ISOs at the solar distance $n$ as
\begin{equation}
n=n_0\left(\frac{D}{D_0}\right)^{-\alpha}. \label{eq:cnd}
\end{equation}
In fact, the power-law parameter $\alpha$ is still unknown. 
Therefore, we vary $\alpha$ as a free parameter. 
We assume that $n$ has the same trend as equation (\ref{eq:galactic_n_ISOs}). 

\subsection{ISO accretion rate}

The key to estimating the metal pollution of \p survivors is how many chances ISOs can approach to the surface of \p survivors. 
We express the ISO accretion rate in the considering epoch $\Delta t_{\rm ISO}$ as
\begin{eqnarray}
\dNaccz = \frac{1}{{\Delta t_{\rm ISO}}}\int_{\Delta t_{\rm ISO}} f n_0(R(t)) \sigma |v(t)-V_{\rm circ}(R(t))| dt, \label{eq:dnaccz_cosmo}
\end{eqnarray}
where $f$ is set to be 1 or 0 in each timestep depending on whether the \p survivor travels an ISO-rich region or not, and $|v(t)-V_{\rm circ}(R(t))|$ is the relative velocity between the \p survivor and an ISO. 
Considering gravitational focusing by the \p star, the cross-section $\sigma$ can be expressed as
\begin{eqnarray}
  \sigma &= \pi r_{*}^2 \left( 1 + \frac{2 G M_{*}}{r_{*} |v(t)-V_{\rm circ}(R(t))|^2} \right), \label{eq:cross-section}
\end{eqnarray}
where $r_*$ and $M_*$ are respectively the radius and the mass of the \p survivor, and $G$ is the gravitational constant. 
When we adopt the solar radius and mass as $r_*$ and $M_*$, the cross-section is written as a function of the relative velocity. 
The stellar mass dependence of $M_*/r_*$ is weak for low-mass main-sequence stars \citep[e.g.,][]{Pagel1997}.
Using the cosmological simulation, we calculate $n_0(R)\sigma |v(t)-V_{\rm circ}(R(t))|$ in each time step monitoring the position of the \p survivor whether or not in the Galactic disc. 

\citet{Tanikawa2018} adopted a constant relative velocity of $310$~$\kms$, which is multiplied by $\sqrt{2}$ to the circular velocity of the Galactic disc. 
The value $f$ was also a constant parameter of $0.032$ that was calculated by the assuming orbit.
Then, \citet{Tanikawa2018} obtained the accretion rate of $1.4\times10^{-4}\, {\rm  yr^{-1}}$. 
It means the number of chances that an ISO accretes to a \p survivor is $1.4\times 10^5$ times per 1~Gyr. 

An ISO accretion rate in mass can be written as
\begin{eqnarray}
  \dMacc = \int_{\Dmax}^{\Dmin} \left\{ \frac{d\dNaccz}{dD}
  \left[ m_0 \left( \frac{D}{D_0} \right)^3 \right] \right\} dD, \label{eq:dmacc1}
\end{eqnarray}
where $m_0$ is the mass of an ISO whose radius is $D_0$. 
We adopt $D_0\sim100\,{\rm m}$ from the observation of `Oumuamua \citep{Do2018}. 
We assume the mass density of an ISO is $3\,{\rm g\,cm^{-3}}$, which is a typical value of asteroids in the solar system \citep{Carry2012}. 
Then we obtain $m_0=1.4\times 10^{13}\, {\rm g}$.

$D_{\rm min}$ is the minimum radius of an ISO, which can collide to the surface of \p survivors. 
When a smaller ISO approaches a \p survivor, it would be radiated and sublimated by the \p survivor. 
Once the ISO is sublimated, the debris would be blown away by stellar wind and would not accrete to the \p survivor. 
According to \citet{Tanikawa2018}, $D_{\rm min}$ is estimated to be $3$~km considering the sublimation effect. 
$\Dmax$ is the maximum radius of ISOs, which can collide with a \p survivor once at least. 
It depends on the accretion rate of ISOs, which we obtain for each \p survivor from our cosmological simulation. 

We can analytically integrate equation (\ref{eq:dmacc1})
\begin{eqnarray}
&\dMacc = \dMaccz \nonumber \\
  &\times \left\{
  \begin{array}{lc}
    \displaystyle \frac{\alpha}{\alpha-3} \left[ \left(
      \frac{\Dmin}{D_0} \right)^{-\alpha+3} - \left( \frac{\Dmax}{D_0}
      \right)^{-\alpha+3} \right] & (\alpha > 3), \\
    \displaystyle \alpha \left[ \log(\Dmax) - \log(\Dmin) \right] &
    (\alpha = 3), \\
    \displaystyle \frac{\alpha}{3-\alpha} \left[ \left(
      \frac{\Dmax}{D_0} \right)^{3-\alpha} - \left( \frac{\Dmin}{D_0}
      \right)^{3-\alpha} \right] & (\alpha < 3),
  \end{array}
\right. \label{eq:dmacc2} 
\end{eqnarray}
where
\begin{eqnarray}
  \dMaccz &= m_0 \dNaccz. \label{eq:dmaccz}
\end{eqnarray}

To calculate $\dot{M}_{\rm acc}$, we here give $\Dmax$ used in equation (\ref{eq:dmacc1}) and (\ref{eq:dmacc2}). 
We define it as the ISO collides with a \p survivor once at least in the considering epoch $\Delta_{\rm ISO}$: 
\begin{eqnarray}
\dNacc \Delta t_{\rm ISO}&\sim & 1. 
\end{eqnarray}
Equation (\ref{eq:galactic_n_ISOs}), (\ref{eq:cnd}) and (\ref{eq:dmaccz}) can be combined to eliminate $\Dmax$: 
\begin{eqnarray}
\Dmax = D_0\left(\dNaccz \Delta t_{\rm ISO}\right)^{1/\alpha}. 
\end{eqnarray}
We obtain $\Dmax$ for each \p survivor from our cosmological simulation.

\subsection{Metal enrichment of \p survivors}

We estimate the metallicity of the surface of \p survivors for three stellar masses ($0.7 \MO$, $0.75 \MO$ and $0.8 \MO$) varying the power-law parameter $\alpha$.
Calculation of the amount of metal supplied by an ISO accretion is completely following \citet{Tanikawa2018}.
We here put a brief description of the method, and \citet{Tanikawa2018} give further details.

We consider surface pollution of \p survivors. 
Accreting metals are mixed only within their convection zone under their surface. 
\p stars are thought to be born after $<1$~Gyr of the Big Bang. 
Their ages are therefore $>12$~Gyr if such stars are observed in the MW. 
According to \citet{Richard2002}, metal-poor stars with the stellar masses of $0.8 \MO$ survive for $>12$~Gyr. 

\citet{Richard2002} calculated stellar evolution and obtained the mass above the base of the surface convection zone in each stellar mass. 
We here adopt their calculation for cases of $0.7 \MO$, $0.75 \MO$ and $0.8 \MO$.
In the case of $0.7 \MO$ and $0.75 \MO$ stars, the mass fraction of convection zones are respectively approximately $10^{-2}$ and $10^{-2.5}$ at $5$~Gyr ago, and these values are slightly decreasing with time. 
In the case of $0.8 \MO$ stars, the fraction rapidly decreases with time from $10^{-3.5}$ at $\sim 5$~Gyr to $10^{-6}$ at the last $\sim 1$~Gyr. 
It means that the efficiency of metal pollution is boosted at the end of their lifetimes. 

We calculate metallicity of a \p survivor as
\begin{equation}
{\rm [Fe/H]}\sim {\rm log_{10}}\left(\frac{1}{f_{\rm conv}}\frac{\dot{M}_{\rm acc}\Delta t_{\rm pol}}{M_* Z_{\odot}}\right), 
\end{equation}
where $\Delta t_{\rm pol}$ is the duration of metal pollution.
The mass fraction of metals in the Sun $Z_{\odot}$ is set to be $0.014$ \citep{Asplund2009}. 
As following the description above, the mass fraction of convection zones $f_{\rm conv}$ is respectively set to $10^{-2}$ and $10^{-2.5}$ for $0.7 \MO$ and $0.75 \MO$ stars, and we adopt $\Delta t_{\rm pol}=\Delta t_{\rm ISO}$. 
Only for $0.8 \MO$ stars, we consider the time dependence of $f_{\rm conv}$. 
We calculate the metallicity setting $f_{\rm conv}=10^{-3.5}$ for 4~Gyr and  $f_{\rm conv}=10^{-6}$ for the last 1~Gyr, which corresponds to $0.2 \Delta t_{\rm ISO}$.

\section{Numerical Results}
\label{sec:ISOacc}
\subsection{ISO accretion rate}

\begin{figure}
 \begin{center}
   \includegraphics[width=80mm]{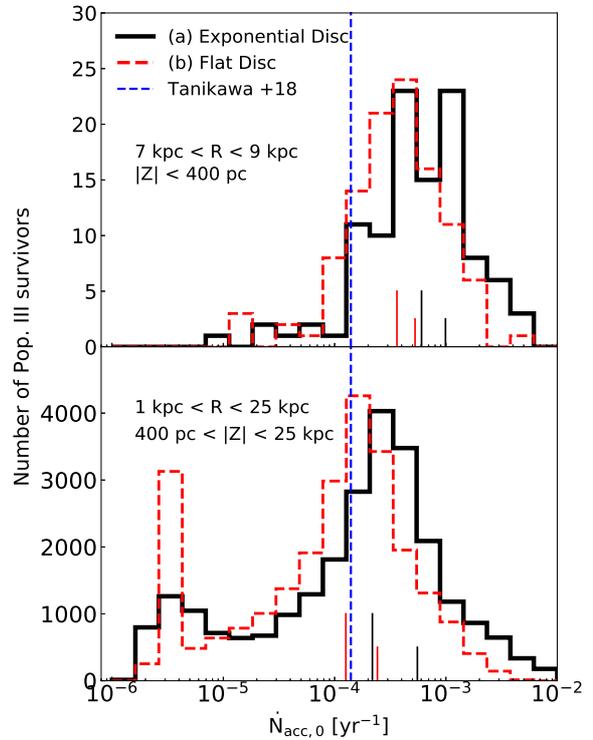}
 \end{center}
  \caption{Top panel: ISO accretion rate onto \p survivors, which are distributed in the Galactocentric distance of $7\, {\rm kpc}<R<9\, {\rm kpc}$ and in the Galactic plane $|Z|<400\, {\rm pc}$ at $z=0$. 
Black solid and red dashed lines show the cases of (a) the exponential disc model and (b) the flat disc model, respectively. 
The blue dashed line indicates the analytical estimation by \citet{Tanikawa2018}. 
Higher and lower bars show the median and average accretion rate, respectively. 
Bottom panel: ISO accretion rate onto \p survivors, which are distributed in the inner galactic halo and outside of the disc ($1\, {\rm kpc}<R<25\, {\rm kpc}$ and $400\, {\rm pc}<|Z|<25\, {\rm kpc}$) at $z=0$. 
Higher and lower bars respectively show the median and average accretion rate except for metal-free samples.  
  }\label{fig:label_nacc}
\end{figure}

Figure~\ref{fig:label_nacc} shows the distribution of the ISO accretion rate $\dNaccz$ onto \p survivors in the last $5$~Gyr ($\equiv \Delta t_{\rm ISO}$). 
We select the \p survivors distributed in the Galactocentric distance of $7\, {\rm kpc} <R<9\, {\rm kpc}$ and in the Galactic plane ($|Z|< 400$~pc) at $z=0$
The median and average accretion rates are showed with higher and lower bars, respectively. 
The typical value of $\dNaccz$ is around $10^{-3}$, which is approximately one order of magnitude greater than the analytical estimation by \citet{Tanikawa2018} as shown with the blue dashed line. 
Even in the case of the flat disc model, which is mimicking \citet{Tanikawa2018}, is higher by a factor of five than the analytical estimation. 
Such a high accretion rate of ISOs onto \p survivors is first revealed by the use of cosmological simulation. 
Orbiting within the Galactic disc and/or similar trail of a \p survivor to the galactic rotation seem to be the causes of greater $\dNaccz$. 
The accretion rate in the case of the exponential disc is roughly higher by a factor of two than that of the flat disc model. 
Since orbiting time in the Galactic disc in each model is the same, it means that most \p survivors have more chance to pass through inner $8\, {\rm kpc}$ than $8\, {\rm kpc}<R<10\, {\rm kpc}$. 
In fact, some \p survivors with higher accretion rate approach Galactic centre in their orbital motion. 
We hereafter consider only the case of the exponential disc model. 
Most noticeably, the survivors have more chances to get metal polluted by ISOs estimated so far. 
If $\dNaccz$ is $10^{-3}\, {\rm yr}^{-1}$, the number of collisions is estimated to $5\times10^6$ times in $5$~Gyr. 

The bottom panel of Fig.~\ref{fig:label_nacc} shows the ISO accretion rates onto \p survivors, which are distributed in the galactic inner halo ($1\, {\rm kpc}<R<25\, {\rm kpc}$ and $400\, {\rm pc}<|Z|<25\, {\rm kpc}$) at $z=0$. 
The samples are less polluted by ISOs floating in the Galactic disc compared with those of the solar neighbourhood. 
Under the condition of the ISO distribution, $\sim$80\% of \p survivors experienced disc-crossing at least once in the last 5~Gyr. 
The median value of the ISO accretion rate for the surface-polluted samples is the same degree estimated by \citet{Tanikawa2018}. 
The median surface metallicity of \p survivors is smaller than that of the solar neighbourhood samples by $\sim 0.5-1$~dex in Fig.~\ref{fig:feh_dist}. 
Well-measured metal-poor stars tend to be located in the solar neighbourhood \citep[e.g.,][]{Sestito2019}; therefore we mainly focus on the samples that are located in the solar neighbourhood.

\subsection{Metal enrichment of \p survivors}
\label{sec:metal_enrichment_local}

\begin{figure}
 \begin{center}
    \includegraphics[width=80mm]{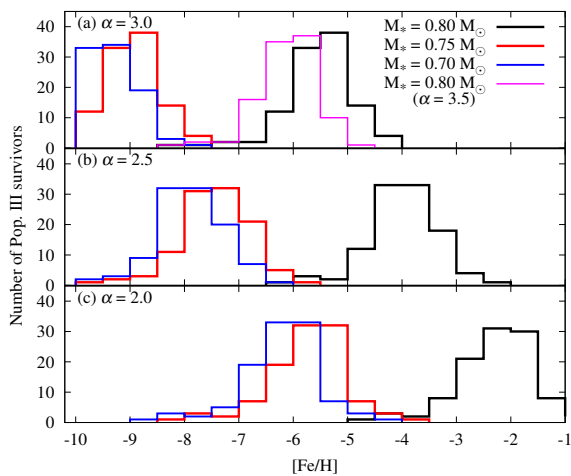}
 \end{center}
  \caption{Number distribution of \p survivors as a function of the surface metallicity [Fe/H] with (a) $\alpha=3.0$, (b) $\alpha=2.5$ and (c) $\alpha=2.0$. 
The difference of each line is the adopted stellar mass of \p survivors of $M_*=0.80 \MO$ (black line), $M_*=0.75 \MO$ (red line) and $M_*=0.70 \MO$ (blue line). 
In panel (a), we additionally plot the case of $(M_*, \alpha)=(0.80 \MO, 3.5)$ with magenta line. 
We select the survivors whose current loci are in the Galactocentric distance of $7\, {\rm kpc}<R<9\, {\rm kpc}$ and in the Galactic disc ($|Z|<400$~pc).
  }\label{fig:feh_dist}
\end{figure}

In Figure~\ref{fig:feh_dist}, we plot the number distribution of \p survivors in solar neighbourhood as a function of [Fe/H].
ISO pollution strongly depends on power-law $\alpha$. 
Typical surface metallicity in all stellar mass models decreases with increasing $\alpha$. 

Surprisingly, in the case of $0.8 \MO$ stars and $\alpha=2.0$, the typical \p survivors are much more metal-rich than the so far observed most metal-poor stars. 
Even in the cases of $0.7$ and $0.75 \MO$ stars, the typical surface metallicity are around [Fe/H]$=-6 \sim -5$. 
The metallicity profile is highly depending on the power-law parameter $\alpha$. 
It should be noted that the metal pollution is caused by cumulative ISO accretion. 

One possibility to distinguish between originally metal-poor stars and metal-polluted \p stars is to see the metal abundance pattern. 
That is because the metal abundance pattern of metal-enriched \p stars is originated in ISOs. 
We expect that the metal abundance pattern is similar to the solar metal abundance pattern. 

\begin{figure}
 \begin{center}
  \includegraphics[width=80mm]{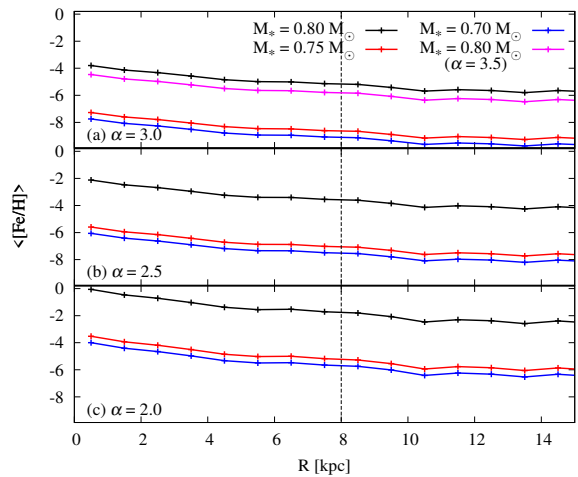}
 \end{center}
  \caption{Average [Fe/H] distribution of \p survivors as a function of the Galactocentric radius $R$ with (a) $\alpha=3.0$, (b) $\alpha=2.5$ and (c) $\alpha=2.0$. 
The difference of each line is the adopted stellar mass of \p survivors of $M_*=0.80 \MO$ (black line), $M_*=0.75 \MO$ (red line) and $M_*=0.70 \MO$ (blue line). 
The magenta line in panel (a) shows the case of $(M_*, \alpha)=(0.80 \MO, 3.5)$. 
We select the survivors whose current loci are in the Galactic disc ($|Z|<400$~pc). 
The vertical dashed line corresponds to the Galactocentric solar distance. 
 }\label{fig:feh_global}
\end{figure}

Figure~\ref{fig:feh_global} shows the average [Fe/H] distribution of \p survivors as a function of the Galactocentric radius $R$. 
To plot this figure, we select the survivors whose current loci are in the Galactic disc ($|Z|<400$~pc). 
From the plot, we obtain a metallicity gradient for typical metallicity of surface metal-polluted \p survivors. 
The average metallicity decreases in all models with increasing the Galactocentric radius. 
When we look at \p survivors near the Galactic centre, they would be polluted typically [Fe/H]$\gtrsim -2$ if $\alpha\lesssim 2.5$. 
We may not be able to distinguish between such highly polluted \p survivors and originally metal-poor stars. 
In each line, there is a break at $R=10$~kpc due to the assuming ISO distribution. 
The \p survivors are located at outer the edge of the Galactic disc at $z=0$. 
This situation causes the break in the profile around the edge. 

\section{Discussion}
\label{sec:Discussion}

\subsection{Constraint on ISO pollution}
\label{sec:ConstraintOnAlpha}

ISOs should pollute not only \p survivors but also originally metal-poor stars, which are born from metal-enriched gas. 
In other words, ISO pollution prohibits the presence of stars with [Fe/H] less than a certain value dependent on $\alpha$. 
Thus, we can constrain on the lower limit of the power $\alpha$ (or the upper limit of ISO pollution) from the presence of metal-poor stars discovered so far. 

For this purpose, we pick up metal-poor stars from the SAGA database \citep{Suda2008}. 
We confine metal-poor stars to those in the MW. 
We can use only main-sequence metal-poor stars. 
Since post main-sequence stars, such as red-giant stars, have massive convection zones, ISO pollution should be diluted by several orders of magnitude and should be erased. 
We adopt the definition of main-sequence stars in the SAGA database. 
We have to estimate the mass and convection zones of metal-poor stars, on which ISO pollution strongly depends (see Figure~\ref{fig:feh_dist}). For this estimate, we refer to effective temperature ($T_{\rm eff}$) of metal-poor stars. 
We assume stars with $T_{\rm eff} \lesssim 6000$~K, $6000~\mbox{K} \lesssim T_{\rm eff} \lesssim 6400~\mbox{K}$, and $T_{\rm eff} \gtrsim 6400~\mbox{K}$ have $0.7$, $0.75$, and $0.8 \MO$, respectively, based on \cite{Richard2002}. 
Note that this can be applied for stars with [Fe/H] $\lesssim -3.31$. 
Stars with $T_{\rm eff} \lesssim 6400$~K can have $0.8 \MO$, when they are young, several Gyr old. 
We can regard such stars ISO-polluted to the same extent as stars with $0.7$ - $0.75 \MO$, since such stars have as thick a convection zone as stars with $0.7$ and $0.75 \MO$ do. 
Thus, our assumption on the relation between stellar mass and $T_{\rm eff}$ is valid.

First, we search for the most metal-poor stars with $0.8 \MO$, since they should be drastically polluted by ISOs (see
Figure~\ref{fig:feh_dist}). 
We find one star with [Fe/H] $< -3.8$ (SDSS~J2209-0028), and six stars with $-3.8 <$ [Fe/H] $<-3.6$ (HE~0228-4047, HE~0049-3948, BPS~CS22876-032A, SDSS~J214633-003910, HE~2032-5633 and SDSS~J0825+0403), where BPS~CS22876-032A indicates binarity. 
We remark that all the above stars have $6400~\mbox{K} \lesssim T_{\rm eff} \lesssim 6600$~K. 
They can never be present typically if $\alpha \lesssim 2.5$. This is because Figure~\ref{fig:feh_dist}
shows half of \p survivors with $0.8 \MO$ are polluted to [Fe/H] $\gtrsim -4$ in the case of $\alpha=2.5$. 
Thus, the lower limit of $\alpha$ can be $2.5$. 
Nevertheless, the lower limit of $\alpha$ might be $2.0$. A slight fraction of \p survivors can remain [Fe/H] $\lesssim -4$, even if $\alpha = 2.0$. 
To assess whether they are the least polluted end, we need the orbits of the seven stars. 
However, their orbits have not been investigated so far.

Second, we search for metal-poor stars with $0.7$ and $0.75 \MO$. 
HE~1327-2326 can put the strictest constraint on $\alpha$ among these stars. 
Its $T_{\rm eff}$ and [Fe/H] are $6180$~K and $-5.71$, respectively. 
We estimate from its $T_{\rm eff}$ that its mass is $0.75 \MO$, and so its [Fe/H] is the lowest among main-sequence stars with $\lesssim 0.75 \MO$. 
It can never be present typically in the case of $\alpha \lesssim 2.0$, since half of \p survivors with $0.75 \MO$ are polluted to [Fe/H] $\gtrsim -5.5$. Thus, the lower limit of $\alpha$ should be $2.0$. 
This constraint is weaker than metal-poor stars with $0.8 \MO$ put, however. 

In summary, we constrain the lower limit of $\alpha$ as $\alpha \gtrsim 2.0$ conservatively. This is consistent with $\alpha$ of km-size asteroids and comets in the solar system
\citep{Gladman2009,Kenyon2004,Fernndez2012}.

\subsection{Candidates of \p survivors}

We assess whether metal-poor stars so far discovered can be \p survivors polluted by ISOs. 
We choose non- carbon-enhanced metal poor (non-CEMP) and main-sequence stars in the MW galaxy as candidates of polluted \p survivors. 
We exclude CEMP stars from these candidates since we conservatively assume ISO compositions are similar to solar abundance pattern. 
Assuming the solar abundance pattern as the abundance of ISOs might lead non-realistic metal pollution of \p stars. 
A part of asteroids exhibits Fe-rich abundant based on the near-infrared absorption features in the reflected light spectra \citep{Ockert-Bell2010}. 
On the other hand, an elemental abundance of directly collected cometary dust of the comet 67P/Churyumov-Gerasimenko was measured \citep{Bardyn2017}, and it showed C/Fe in number as $\sim 19$. 
Considering C/Fe of the Sun \citep{Lodders2010}, [C/Fe] of the commet $\sim \, {\rm log}(19/8.5)\sim 0.35$. 
Although observations of the surface elemental abundance of them show different patterns with solar abundance, the inner elemental abundances of comets and asteroids are still less knowledgeable. 
Even if individual ISOs have peculiar abundance pattern, their peculiarities should be smoothed out by collisions million times during several Gyrs (see Figure~\ref{fig:label_nacc}). 
We, therefore, assume the solar abundance for ISOs as a safer working hypothesis. 
We focus on only main-sequence stars. 
Post main-sequence stars have thick convection zone. We pick up these candidates from the
SAGA database \citep{Suda2008} again. We adopt the definition of non-CEMP and main-sequence stars in the SAGA database.

Whether a metal-poor star is a \p candidate strongly depends on the power $\alpha$. 
Thus, we show \p candidates in each $\alpha$. 
In the case of $\alpha \gtrsim 3.0$, there is no \p candidate.
In the case of $\alpha \sim 2.5$, $0.80 \MO$ stars shown in
\S~\ref{sec:ConstraintOnAlpha} (HE~0228-4047, HE~0049-3948, BPS~CS22876-032A, SDSS~J214633-003910, HE~2032-5633 and SDSS~J0825+0403) are the most promising candidates of \p survivors. 
Note that SDSS~J2209-0028 is a CEMP-s star.

In the case of $\alpha \sim 2.0$, non-CEMP and main-sequence stars with $T_{\rm eff} \sim 6500$~K ($0.80 \MO$) and [Fe/H] $\lesssim -3$ are candidates of \p survivors, which include the six stars. 
Moreover, there is a \p candidate with $\lesssim 0.75 \MO$: SDSS~J164234+443004 ($T_{\rm eff}=6280$~K and [Fe~I/H]$=-4.05$). 
From its $T_{\rm eff}$, it should have $0.75 \MO$. 
Thus, it has to be the most polluted end as seen in Figure~\ref{fig:feh_dist}. Nevertheless, it should be a \p candidate in the following reason. 
From our analysis, we find the most polluted end has low altitude from the Galactic disc ($\lesssim 4$~kpc), and small perigalacticon ($\sim 2$~kpc). 
SDSS~J164234+443004 has such an orbit \citep{Sestito2019}. 

There are several candidates with [Fe/H] $\lesssim -4$: SDSS~J102915+172927 ($T_{\rm eff}=5811$~K and [Fe~I/H]$=-4.71$),
HE~2239-5019 ($T_{\rm eff}=6100$~K and [Fe~I/H]$=-4.15$) and SDSS~J144256-001542 ($T_{\rm eff}=5850$~K and [Fe~I/H]$=-4.09$). 
However, they must not be the most polluted end from their orbits obtained from \cite{Sestito2019}, and so must not be ISO-polluted \p survivors. 
Although SDSS~J102915+172927 has low altitude from the Galactic disc ($\sim 2$~kpc), its perigalacticon is $\sim 8$~kpc. 
HE~2239-5019 and SDSS~J144256-001542 have high altitudes from the Galactic disc to several $10$~kpc. 
Although metal-poor stars with $-4 \lesssim $ [Fe/H] $\lesssim -3.5$ should be assessed, their orbits have not been in the literature.

In summary, there are six candidates in the case of $\alpha \sim 2.5$: HE~0228-4047, HE~0049-3948, BPS~CS22876-032A, SDSS~J214633-003910, HE~2032-5633 and SDSS~J0825+0403. 
In the case of $\alpha \sim 2.0$, there are many candidates with $0.80 \MO$ including the above six stars, and moreover there is one candidate with $\lesssim 0.75 \MO$: SDSS~J164234+443004. Note that we exclude a CEMP star SDSS~J1035+0641 from these candidates, although \cite{Tanikawa2018} have said it can be a \p candidate. 

\subsection{Strategy to discover \p survivors}

We cannot distinguish between polluted \p survivors and originally metal-poor stars since the abundance pattern of ISOs has
been unknown. 
Thus, we should consider a method to search for unpolluted \p survivors in order to discover them.

We should avoid the following two types of stars. The first type is stars with $0.8\MO$ (or $T_{\rm eff} \sim 6500$~K). We can see from Figure~\ref{fig:feh_global} that these stars are typically polluted to
[Fe/H] $\sim -4$ over the MW if $\alpha \lesssim 2.5$. 
The second type is stars with $\lesssim 0.75\MO$ (or $T_{\rm eff} \lesssim 6400$~K) in the Galactic disc, and at the Galactic centre. 
They can be polluted to [Fe/H] $\sim -4$ in the case of $\alpha \sim 2.0$ (see Figures~\ref{fig:feh_dist} and \ref{fig:feh_global}).

In short, we may possibly discover unpolluted \p survivors, surveying stars with $T_{\rm eff} \lesssim 6400$~K at the Galactocentric radii of $\gtrsim 15$~kpc. 
At those radii, these stars should be typically polluted only to [Fe/H] $\sim -6$ even if $\alpha \sim 2.0$ (see Figure~\ref{fig:feh_global}).

The number estimation of low-mass \p stars in the MW is another important information to find low-mass \p stars \citep{Hartwig2015,Ishiyama2016,Griffen2018,Magg2018,Magg2019}. 
However, the number highly depends on the assumption of the low-mass cutoff of the initial mass function of \p stars. 
Moreover, low-mass \p stars are assumed to be still metal-free stars in the MW in previous studies.  
Considering the observational bias of metal-poor stars, \citet{Magg2019} concluded that a more significant number of metal-free stars should be observed unless reducing the star formation of low-mass \p stars or external pollution of metal-free stars. 
Our results suggest that the apparent number of metal-free stars get smaller due to the surface ISO metal pollution in the solar neighbourhood.

\section{Summary}
\label{sec:summary}

In this work, we investigated metal pollution of \p survivors via ISOs floating in the Galactic interstellar medium. 
Using a high-resolution cosmological $N$-body simulation, we calculated the accretion rate in each time step monitoring the position of the \p survivor whether or not in the Galactic disc. 
To take realistic orbits of \p survivors based on the cosmological framework into calculating the accretion rate, we obtained that the typical accretion rate is one order of magnitude greater than estimated by the previous analytical study \citep{Tanikawa2018}.
When we give a cumulative number density of ISOs as described in \S \ref{sec:ISONumberDensity}, we can obtain that the number of chances of ISO($> 100$~m) collisions is typically $5\times10^6$ times in the last $5$~Gyr for \p survivors near solar neighbourhood. 
 
When we adopt the ISO cumulative number distribution expressed with a power-law parameter $\alpha$, $0.80 \MO$ stars should be typically polluted [Fe/H]$\sim -2, -4$ and $-5.5$ for the case of $\alpha=2.0, 2.5$ and  $3.0$, respectively. 
We can constrain on the lower limit of the power $\alpha$ as $\alpha \gtrsim 2.0$ from the presence of so far observed metal-poor stars. 
The constraint is consistent with $\alpha$ of km-size asteroids and comets in the solar system.
Furthermore, we provide six candidates as the ISO metal-enriched \p stars in the case of $\alpha \sim 2.5$.
Owing to the Gaia photometric and astrometric observation, highly-accurate orbital information of each star is getting available \citep{GaiaDR22018}. 
Most earlier studies have considered that the observed metallicity of stars is unchanged from their birth. 
We again note that metal-poor stars so far discovered are possible to be metal-free \p stars on birth and are later metal-polluted by ISOs.

\section*{Acknowledgements}

TK is grateful to Yu Morinaga for providing fruitful suggestions about the treatment of the merger tree. 
We thank an anonymous referee for the constructive comments. 
Numerical calculations were partially carried out on Aterui supercomputer at Center for Computational Astrophysics, CfCA, of National Astronomical Observatory of Japan and the K computer at the Riken Institute for Computational Science (Proposal numbers hp150226 and hp160212). 
This research has been supported in part by MEXT program for the Development and Improvement for the Next Generation Ultra High-Speed Computer System under its Subsidies for Operating the Specific Advanced Large Research Facilities, and by Grants-in-Aid for Scientific Research (16K17656, 17H01101, 17H04828, 17H06360, 18H04337 and 19K03907) from the Japan Society for the Promotion of Science. 
We thank the SAGA developers for making up such a useful database.




\bibliographystyle{mn2e}
\bibliography{ms}







\bsp	
\label{lastpage}
\end{document}